\documentstyle[prl,aps,preprint,tighten,floats,epsf,rotate]{revtex}

\newbox\rotbox

\begin{document}
\draft
\setcounter{page}{0}
%
\title{Chirality and Reliability of Baryon QCD Sum Rules\thanks
{This work is supported in part by funds provided by the U.S.
Department of Energy (D.O.E.) under cooperative 
research agreement \#DF-FC02-94ER40818.}}

\author{Xuemin Jin\footnote%
{Email address: {\bf jin@ctpa02.mit.edu}}
 and Jian Tang\footnote%
{Email address: {\bf jtang@ctpa02.mit.edu}}}

\address{Center for Theoretical Physics \\
Laboratory for Nuclear Science \\
and Department of Physics \\
Massachusetts Institute of Technology\\
Cambridge, Massachusetts 02139, USA \\}

\date{MIT-CTP-2601 ~~~~~ hep-th/9701230 {~~~~~} January 1997}
\maketitle

\thispagestyle{empty}

\begin{abstract}

The QCD sum-rule method has been widely used in studying various baryon 
properties. For a given problem, there are usually more than one sum rules 
and they do not work equally well. In this paper, we point out that chirality 
plays an important role in determining the reliability of a baryon sum rule. 
The contributions of positive- and negative-parity excited baryon states partially 
cancel each other in the chiral-odd sum rules, but add up in the chiral-even sum rules. 
As such, the chiral-odd sum rules are generally more reliable than the chiral-even sum rules.
This allows one to identify the more reliable sum rules and use them in extracting 
the ground-state baryon property of interest. This is illustrated in an explicit example.

\end{abstract}

\vspace*{\fill}
\begin{center}
Submitted to: {\it Physical Review D}
\end{center}


\newpage

The QCD sum-rule approach~\cite{svz79} has been used extensively in 
extracting baryon observables~\cite{reinders85,spires}. It is found in these 
studies that for a given problem there are more than one sum rules and they do 
not work equally well. In particular, some sum rules work well while the 
others may fail. This pattern has been seen in both the baryon mass sum rules and 
the sum rules for other baryon properties~\cite{spires}. 
(Here we consider the baryons containing only light flavors: up, down, and strange.) 

In this paper, we point out that chirality plays an important role 
in determining the reliability of baryon QCD sum rules. In the ``chiral-odd''
sum rules where chiral-odd operators dominate, the contributions 
of positive- and negative-parity excited baryon states partially cancel 
each other, which significantly reduces the excited-state contamination.
On the other hand, the ``chiral-even'' sum rules where chiral-even 
operators dominate, suffer from large excited-state contamination because of
the addition of the contributions from positive- and negative-parity excited
states. As such, the chiral-odd sum rules are generally more reliable 
than the chiral-even sum rules. This allows one to identify the more
reliable sum rules in a given problem and use them in extracting
the ground-state baryon property of interest. We will illustrate this point 
in an explicit example.

To keep our discussion succinct, we will use the octet baryon mass sum rules to 
explain the idea, which applies to general baryon sum rules based on two-point 
and three-point correlation functions. The baryon mass sum rules usually study 
the correlation function 
\begin{equation}
\Pi(q) \equiv i \int d^4x e^{iq\cdot x}
\langle 0|T[\eta_B(x) \overline{\eta}_B(0)]|0\rangle\ ,
\label{corr-gen}
\end{equation}
where $\eta_B$ is a baryon interpolating field constructed from
local QCD operators, carrying the quantum numbers
of the baryon of interest. For the octet baryons, there are two 
independent interpolating fields which contain only quark fields
with no derivatives and couple to spin-1/2 states only~\cite{ioffe81,chung82}. 
The general expression of the proton interpolating field is 
\begin{equation}
\eta_p(x) = \epsilon_{abc}
\left\{\left[u^{aT}(x)C\gamma_5 d^b(x)\right] u^c(x)
+ \beta \left[u^{aT}(x)C d^b(x)\right]\gamma_5 u^c(x)
\right\}\ ,
\label{eta-p}
\end{equation}
where $u(x)$ and $d(x)$ stand for up and down quark fields, $a$, $b$, and $c$
are the color indices, $C = -C^T$ is the charge conjugation matrix, and 
$\beta$ is an arbitrary real parameter. The corresponding interpolating fields 
for the other members of the octet can be obtained by applying appropriate 
SU(3) rotations~\cite{reinders85}. The interpolating field advocated by 
Ioffe~\cite{ioffe81} and often found in QCD sum-rule calculations may be 
recovered by setting $\beta = -1$ and multiplying an overall factor of $-2$. 
Lorentz covariance and symmetries then dictate that $\Pi(q)$ has two invariant 
structures 
\begin{equation}
\Pi(q) = \Pi_s(q^2) + \Pi_q(q^2)\,  q_\mu\gamma^\mu\ ,
\label{decomp-gen}
\end{equation}
where $\Pi_s$ and $\Pi_q$ are two invariant functions. So, two
sum rules, one for $\Pi_s$ and one for $\Pi_q$, can be obtained.
In principle, results obtained from these sum rules should be the same. 
In practice, however, one has to truncate the operator product 
expansion (OPE) on the QCD side of the sum rule and use a crude model 
for the excited-state (continuum) contribution on the phenomenological 
side of the sum rule. Thus, one expects that some sum rules are more 
reliable than the others.

One of the key ingredients of the QCD sum-rule approach is the
use of the Borel transform, which introduces an auxiliary
parameter--Borel mass. If a sum rule were perfect, one
would expect that the two sides of the sum rule overlap for
all values of the Borel mass. In practical calculations,
the two sides of the sum rules overlap only in a limited
range of the Borel mass (at best) because of the truncation
of the OPE and the crudity of the continuum model. A common consensus
is that if there exists a wide region of Borel mass where the
two sides of a sum rule match, the sum rule is said to ``work''
and the sum-rule prediction is possible. This, however, needs
to be interpreted carefully.

To maintain the predicative power of the sum-rule approach, the 
phenomenological side of a sum rule is usually parameterized
by a pole describing the ground-state baryon of interest plus
a continuum model accounting for the contribution of all excited 
states. The continuum model approximates the excited-state contributions 
in terms of the perturbative evaluation of the invariant functions,
starting from an effective threshold; it is thus a crude model.
In order to extract the spectral parameters of the pole  
by matching the sum rules, one should work in a 
region of Borel mass where the pole contribution dominates the phenomenological
side. This usually sets an upper bound in the Borel mass space, beyond which  
the excited-state contribution dominates. 
On the other hand, the truncated OPE must be sufficiently convergent 
as to accurately describe the true OPE. This, in practice, sets a lower limit
in the Borel mass space, beyond which higher order terms not present
in the truncated OPE may be significant and important. 

Therefore, we expect a sum rule to work if the two sides of the sum rule match
in a {\it valid} window in Borel mass space where the pole contribution to 
the phenomenological side dominates and the higher order OPE terms
are under control. The relative reliability of a sum rule is thus 
determined by the quality of overlap of the two sides, the size
of valid Borel window, and the relative contribution of the excited
states in the valid Borel window. Such criteria have been advocated 
in Refs.~\cite{leinweber90,leinweber95} and used in various sum rule 
calculations~\cite{leinweber90,leinweber95,jin95,iqbal96,furnstahl96,lee96}. 
Here, we adopt these criteria to determine the relative reliability of a sum rule.

Let us now focus on the phenomenological representation of $\Pi(q)$, 
which can be obtained by inserting a complete set 
of eigenstates with the quantum numbers of the interpolating field.
Since $\eta_B$ couples to not only the positive-parity baryon states
but also the negative-parity baryon states~\cite{chung82}, $\Pi(q)$ can 
be expressed as 
\begin{equation}
\Pi^{\rm phen}(q) = -\lambda_0^2 {q_\mu\gamma^\mu +M_0
\over q^2 -M_0^2} - \sum_{i\neq 0} (\lambda_i^+)^2 \, {q_\mu\gamma^\mu + M_i^+
\over q^2 - (M_i^+)^2} 
- \sum_i (\lambda_i^-)^2 \, {q_\mu\gamma^\mu - M_i^-
\over q^2 - (M_i^-)^2}\ ,
\label{corr-phen}
\end{equation}
in the discrete-state approximation (i.e., neglecting the widths of 
baryon states, which is usually assumed in the QCD sum-rule approach),
where ``0'' denotes the ground state baryon of interest with mass $M_0$, and
``$+$''/``$-$'' the positive/negative parity excited baryon
state with mass $M_i^+/M_i^-$. Here we have omitted the infinitesimal 
as we are only concerned with large and space-like $q^2$. The $\lambda^\pm_i$
($\lambda_0\equiv \lambda^+_0$) stands for the coupling strength of $\eta_B$ 
to the physical baryon state :
\begin{equation}
\langle 0|\eta_B|i,q,s,+\rangle = \lambda_i^+ U(q,s)\ ,
\hspace*{1cm}
\langle 0|\eta_B|i,q,s,-\rangle = \lambda_i^- \gamma_5 U(q,s)\ ,
\label{lambda-def}
\end{equation}
with $U(q,s)$ the baryon Dirac spinor. After the Borel transform, one 
obtains from Eq.~(\ref{corr-phen})
\begin{eqnarray}
\Pi^{\rm phen}_s(M^2) &=& \lambda_0^2\, M_0\, e^{-M_0^2/M^2}
+\sum_{i\neq 0} (\lambda_i^+)^2 M_i^+ e^{-(M_i^+)^2/M^2}
-\sum_i(\lambda_i^-)^2 M_i^- e^{-(M_i^-)^2/M^2}\ ,
\label{pis-phen-borel}
\\*[7.2pt]
\Pi^{\rm phen}_q(M^2) &=& \lambda_0^2\, e^{-M_0^2/M^2}
+\sum_{i\neq 0} (\lambda_i^+)^2 e^{-(M_i^+)^2/M^2}
+\sum_i (\lambda_i^-)^2  e^{-(M_i^-)^2/M^2}\ ,
\label{piq-phen-borel}
\end{eqnarray}
where $M$ is the Borel mass. 
We observe that the positive- and negative-parity excited-state 
contributions partially cancel each other in $\Pi^{\rm phen}_s(M^2)$,
but add up in $\Pi^{\rm phen}_q(M^2)$. This indicates that the continuum 
contamination is significantly smaller in the sum rule for $\Pi_s$ than in 
that for $\Pi_q$. 

This feature attributes to chirality. If there were no spontaneous and explicit 
chiral symmetry breaking, the chirality of quarks could not be changed during 
propagation, and there would be chiral doubling in the baryon states and 
$\Pi^{\rm phen}_s(M^2) =0$. In nature, the chiral symmetry is violated by 
QCD vacuum and (small) current quark masses, which can cause quark chirality
flipping and hence the shift of the positive-parity states relative to the 
negative-parity states. The two invariant functions $\Pi_s$ and $\Pi_q$ correspond 
to the chirality-changing and chirality-conserving parts of $\Pi$, respectively. 
The chirality-conserving part $\Pi^{\rm phen}_q(M^2)\neq 0$ even if there were 
no chiral symmetry violation. This gives rise to extra background noises in the 
sum rule for $\Pi_q$. Thus, the QCD side of the sum rule for $\Pi_s$ must be dominated 
by chiral-odd operators (e.g., $\overline{q}q$, $\overline{q} g_s\sigma\cdot G q$) as 
the current quark masses are small compared to the baryon masses and the QCD side of 
the sum rule for $\Pi_q$ must be dominated by chiral-even operators (e.g., 1, $G^2$). 
We define ``chiral-odd'' sum rules as those where the chiral-odd operators 
dominate and ``chiral-even'' sum rules where the chiral-even operators dominate.
Therefore, the chiral-odd sum rules are generally more reliable than the 
chiral-even sum rules.

As an explicit example, we consider the sum rules for the proton matrix element 
of $H = \langle p|\overline{u}u-\overline{d}d|p\rangle/2M_N$, which can be  
obtained from the linear response of $\Pi$ to a constant external isovector-scalar
field~\cite{jin95a,jin95b}. In the formalism suggested in Ref.~\cite{jin96}, 
the two sum rules given in Ref.~\cite{jin95b} can be rewritten as:
\vspace*{1.5cm}
\begin{eqnarray}
& &{c_1\over 32} a \left(M_N^2 M^2 -M^4\right) L^{-4/9}
-{c_2\over 48}m_0^2 a M_N^2 L^{-24/27}
\nonumber
\\*[7.2pt]
& &\hspace*{1.0cm}
+{c_3\over 12}\chi a^2 M_N^2 L^{4/9}
-{c_3\chi\over 48} \left(1+{M_N^2\over M^2}\right)m_0^2 a^2 L^{-2/27}
\nonumber
\\*[7.2pt]
& &\hspace*{1.0cm}
= 2 H \widetilde{\lambda}^2_N M_N e^{-M_N^2/M^2}
+{c_1\over 32} a \left(M_N^2 M^2 -
 M^4 \widetilde{E}_1\right)L^{-4/9}
e^{-s_1/M^2}\ ,
\label{iso-odd}
\end{eqnarray}
\begin{eqnarray}
& &-{c_4\over 16}\left(M_N^2 M^6 -3 M^8\right) L^{-8/9}
+{c_4\over 16}\chi a \left(M_N^2 M^4 -2 M^6\right)
\nonumber
\\*[7.2pt]
& &\hspace*{1.0cm}
-{3 c_5\over 16}\chi m_0^2 a \left(M_N^2 M^2 - M^4\right) L^{-14/27}
-{c_6\over 24} a^2 M_N^2
\nonumber
\\*[7.2pt]
& &\hspace*{1.0cm}
= 2 H \widetilde{\lambda}^2_N M_N^2 e^{-M_N^2/M^2}
-{c_4\over 16}\left(M_N^2 M^6 \widetilde{E}_2
 -3 M^8\widetilde{E}_3\right) L^{-8/9}
e^{-s_2/M^2}
\nonumber
\\*[7.2pt]
& &\hspace*{1.5cm}
+{c_4\over 16}\chi a \left(
M_N^2 M^4\widetilde{E}_1
-2 M^6\widetilde{E}_2\right)e^{-s_2/M^2}
\nonumber
\\*[7.2pt]
& &\hspace*{1.5cm}
-{3 c_5\over 16}\chi m_0^2 a \left(M_N^2 M^2 -
 M^4\widetilde{E}_1 \right) L^{-14/27}
e^{-s_2/M^2}\ ,
\label{iso-even}
\end{eqnarray}
where $s_1$ and $s_2$ denote the continuum thresholds,  
$c_1 = 1+10\beta+\beta^2$, $c_2 = 4+7\beta+\beta^2$,
$c_3 = 1-2\beta+\beta^2$, $c_4 = 5+ 2\beta -7\beta^2$, $c_5=1-\beta^2$,
$c_6 = 7+4\beta+\beta^2$, $\widetilde{E}_1 = {s_i\over M^2}+1$,
$\widetilde{E}_2 ={s_i^2\over 2M^4}+{s_i\over M^2} +1$, 
$\widetilde{E}_3= {s_i^3\over 6M^6}+{s_i^2\over 2M^4}+{s_i\over M^2}+1$,
$a= -(2\pi)^2\langle\overline{q}q\rangle$,
$m_0^2 = \langle\overline{q}g_s\sigma\cdot G q\rangle
/\langle\overline{q}q\rangle$, $L =\ln(M/\Lambda_{\rm QCD})/
\ln(\mu/\Lambda_{\rm QCD})$, and $\widetilde{\lambda}_N
= (2\pi)^2 \lambda_N$. For definiteness, we use $\beta = -1.2$~\cite{leinweber95}
and $\chi = 2.0$ GeV$^{-1}$~\cite{jin95a,jin95b}.  Here, we have put the 
continuum contributions on the right-hand sides of the sum rules.

One notices that sum rule (\ref{iso-odd}) is the chiral-odd sum rule, which,
according to the above discussion, is more reliable than the chiral-even sum 
rule (\ref{iso-even}).
This has been emphasized in Refs.~\cite{jin95a,jin95b}. Here, we give a numerical
demonstration. We take the experimental value for the nucleon mass $M_N$ and 
extract $H$ and the continuum thresholds from the sum rules. To this end,
we sample the sum rules in a valid Borel region, where the continuum
contribution is less than $50\%$ of the total phenomenological (right-hand)
side and the contribution of the highest order OPE term is less 
than $10\%$ of the total QCD (left-hand) side. The fit of the two sides
is measured by $\delta(M^2) = |{\rm LHS-RHS}|^2$ averaged over 150 points
evenly spaced within the valid Borel region. 

\begin{figure}[t]
\begin{center}
\epsfysize=15.7truecm
\leavevmode
\setbox\rotbox=\vbox{\epsfbox{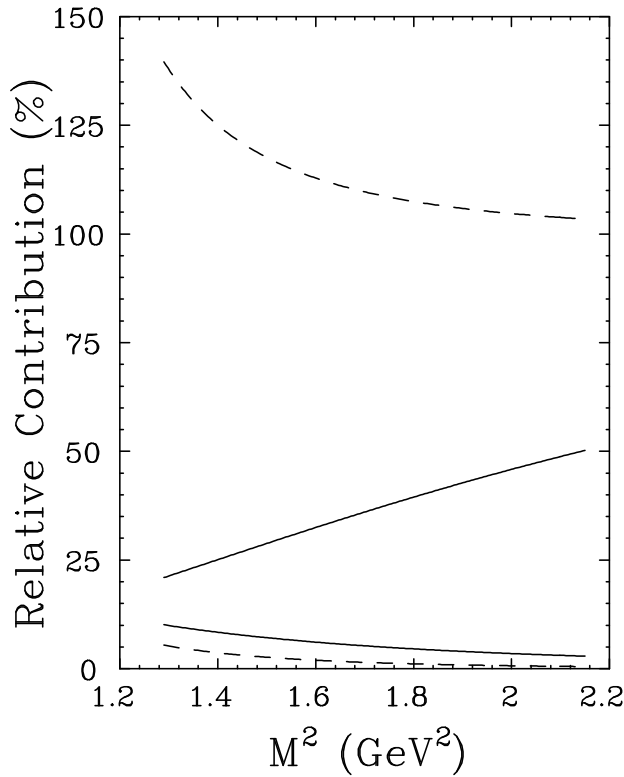}}\rotl\rotbox
\end{center}
\caption{The relative continuum (upper) and highest order OPE term (lower)
contributions in the valid Borel window for the chiral-odd sum rule of 
(\protect{\ref{iso-odd}}), with $H$ and the continuum thresholds extracted
from (\protect{\ref{iso-odd}}). The solid curves are for the chiral-odd sum rule
of (\protect{\ref{iso-odd}}) and the dashed curves for the chiral-even
sum rule of (\protect{\ref{iso-even}}). Note that the relative 
continuum contribution in the chiral-even sum rule are greater than $100\%$.
This results from partial cancelation between the first term and the rest
of the right-hand side of (\protect{\ref{iso-even}}).}
\label{fig-1}
\end{figure}

We find that the two sides of the chiral-odd sum rule (\ref{iso-odd}) 
overlap very well in a large valid Borel region. In Fig.~\ref{fig-1}, 
we have plotted the relative continuum and highest order OPE term 
contributions in the valid Borel window. In contrast, there is no 
valid Borel window for the chiral-even sum rule (\ref{iso-even}). This is 
mainly due to the large continuum contamination. To stress this
point, we have displayed in Fig.~\ref{fig-1} the relative continuum and 
highest order OPE term of the chiral-even sum rule within the valid
Borel window of (\ref{iso-odd}), with the use of $H$ and the continuum 
threshold extracted from the chiral-odd sum rule. One can see that 
the continuum contribution is dominant in the chiral-even sum rule, making 
it impossible to isolate the ground state signal of interest in this 
sum rule. This may also be understood from the different behavior of the 
continuum model in the two sum rules. While the leading Borel
mass dependence is $M^4$ in (\ref{iso-odd}), it is $M^8$ in
(\ref{iso-even}). So, the continuum contribution grows much
more rapidly in (\ref{iso-even}) than in (\ref{iso-odd}) as
$M^2$ increases, implying much larger continuum contamination
in (\ref{iso-even}) than in (\ref{iso-odd}).

Thus, if the OPE is truncated at the order usually considered in the literature, 
the chiral-even sum rules are likely to fail to have a valid Borel window.  
To improve this situation, one may have to include many higher order OPE terms 
in order to compensate for the smaller upper bound in Borel mass space compared 
to the chiral-odd sum rule. However, one usually does not have much control over 
the values of such terms, which makes it difficult to extract any useful information 
about the ground state from the chiral-even sum rules.

In the literature it is often found that both chiral-odd and chiral-even
sum rules are used in extracting baryon properties. In some cases,
certain chiral-even sum rules are identified as the most reliable 
sum rules. This largely relies on the so-called ratio method widely adopted in the 
literature. There, one chooses the continuum threshold (to be the same for two
different sum rules) to make the ratio of two {\it different} sum rules
as flat as possible as a function of the Borel mass. 

However, it is worthwhile emphasizing that the ratio method has serious drawbacks. 
(1) The ratio method does not check the validity and reliability of each 
individual sum rule. It may happen that individual sum rules are not valid 
(i.e., there does not exist a valid Borel window) while their ratio is flat 
as function of the Borel mass. (2) The ratio method may lead 
to misleading results as the continuum contributions are not monitored 
in the ratio. It is obvious that the ratio is always perfectly 
flat for large Borel mass, which is trivial because the continuum 
contribution is dominant in a sum rule and modeled by the perturbative 
evaluation of the OPE. There, however, one should not expect to get any 
reliable information about the ground state. (3) The continuum threshold is 
a phenomenological parameter introduced to parameterize the spectral function. 
Hence, one should treat the continuum threshold as an independent parameter 
(just as the other parameters, masses, coupling strength, ect.) to be extracted 
from the sum rule. Fixing the continuum threshold will introduce 
artificial bias to the extracted baryon properties of interest. 

To summarize, we have pointed out that chirality plays an important role 
in determining the reliability of baryon QCD sum rules. In particular, 
the contributions of positive- and negative-parity excited baryon states 
partially cancel each other in the chiral-odd sum rules, but add up in the 
chiral-even sum rules. Consequently, the continuum contamination is much smaller 
in the former than in the latter. Moreover, there are other 
(relatively) large uncertainties in the chiral-even sum rules arising from 
large perturbative corrections to the Wilson coefficients 
in the OPE, the factorization assumption for chiral-even operators, and the 
onset of nonfactorizable operators in relatively low dimension~\cite{leinweber95}.
Thus, the chiral-odd sum rules are more reliable than the chiral-even sum rules.
The chiral-odd sum rules can be easily identified for a given problem and should 
be used in extracting various baryon properties. On the other hand, without
further improvement, the chiral-even sum rules are likely to fail in most cases and 
hence should be disregarded. Any misuse of the chiral-even sum rules could lead
to inconsistent and misleading results.




\begin{references}
\bibitem{spires}There have been many works studying baryon properties
in the QCD sum-rule approach. These works can be retrieved easily
from the SLAC SPIRES.
\end{references}
\end{document}